# Analysis of Electricity Production Networks Using Mathematical and Simulation Method in Photovoltaic Power Generations

Mohsen MomeniTabar, Zhila Dehdari Ebrahimi, Mohammad Arani,
North Dakota State University, North Dakota State University, University of Arkansas at Little Rock
mohsen.momenitabar@ndsu.edu, zhila.dehdari@ndsu.edu, mxarani@ualr.edu

*Abstract*— currently waste of energy is one of the most important issues, power plants are encountered all over the world. In this study, two mathematical programming models are presented to formulate solar power plants pertinent to the design of a lean manufacturing system. We propose an approach consisting of a combination of mathematical and economic models used for distribution systems of electricity transmission networks. Besides, system's diverse costs and the rate of energy waste during the transmission are considered as two criteria for the comparison. The main approach is to employ simulation method for both models and draw a comparison between the results of simulation and optimization models run by Lingo software. Finally, analyzing the test results lead the system to remarkable cost reduction of country's energy demand in distribution system of electricity transmission networks.

*Index Terms*— Operations Research, Solar Power Plants, Simulation Method, Lean Manufacturing, Photovoltaic Power Generations.

## I. INTRODUCTION

There is an urgent need to accelerate the development of advanced clean energy technologies all over the world in order to decrease the effect of global warming as well as the threatening major changes to the climate caused by fossil fuels. Solar energy is believed to be a kind of renewable, sustainable and endless energy. [1] [2] Solar energy, when used in practice such as lighting, space heating, and water heating, is a proven viable alternative to non-renewable energy sources [3]. Therefore, using solar power plants can be a reasonable replacement for old power plants causing atmospheric pollution on a massive scale. The most important application of solar power plants is electricity production which provides energy for performing different activities as a vital demand for a modern society. To add value to the distributed nature of generated solar electricity, it is important to know the photovoltaics (PV) capacity of the different regions of a city when installing a PV power plant in order to select the feeder with the greatest capacity credit [4], [5], and [6]. The photovoltaic electricity generation in Taiwan had the potential of 36.1 TWh, accounting for 16.3% of the total domestic electricity consumption [7]. The daily average solar radiations of Gaize and Florianopolis were ranged from 14 to 27 MJ/m2 and from 2.46 to 5.72 kWh/m2 respectively [8]–[14] . The annual average solar radiation was obtained between 750 kWh/m2 and 2485 kWh/ m2 for Lake Van Region in Turkey [15]. The rooftop solar photovoltaic potentials were quantified as 885.1 kWh and 10 *TWh* in respectively [16], and [17]. The hybrid photovoltaic/thermal solar system increased the production of electrical and thermal energy by 38% [18]. The cumulative installed photovoltaic power increased to 1.45 GW in the USA, 2 GW in Japan and 5.3 GW in Germany by the end of 2006–2008, respectively [19]. Most of the studies given above generally use the Weibull model, Rayleigh distribution model, maximum entropy theory, artificial neural networks, feature extraction algorithms, and geographical information systems in the stage of analyzing wind and solar power potentials [20]. Besides, the above researches show the importance of PV solar power plants utilized in different countries, there have been different researches to study the performance and financial improvement of various solar power plants and PV generations. Pasumarthi and Sherif built a complete mathematical model and analyzed the experimental and theoretical performances of a demonstration solar chimney power system [21], and [22].

Zhao et al., [23] presented the design optimization of a photovoltaic/thermal (PV/T) system using both non-concentrated and concentrated solar radiations. The system consisted of a photovoltaic (PV) module using a silicon solar cell and a thermal unit based on the direct absorption collector (DAC) concept Shah et al. [24] studied the impact of large scale PV generation on power system oscillation, through which the inter-area oscillation was specially considered. The effect of PV on inter-area mode was investigated in New England–New York test system for different levels of penetrations and operating conditions. The analysis revealed that increased PV penetration could affect the critical inter-area mode detrimentally. Mediavilla et al. [25] studied real PV production from two 100 kWp grid-connected installations located in the same area, both of which experienced the same fluctuations in temperature and radiation.

Although there have been many types of research related to solar power plants and PV generations, notably PV ones, there has hardly ever been a significant attempt to present mathematical programming and simulation methods applied to solar power plants and photovoltaic power generations on buildings as a lean manufacturing method. Thus, in this paper, it has been aimed to study solar power plants and photovoltaic generations on buildings. The former is installed out of a city where different parameters are considered including the cost of production, the costs of power transition, the additional costs of power, set up costs for every solar power plant, etc. However, the latter is a lean manufacturing model consisting of solar panels for every building in the city separately. Similarly, different aspects of the model are also presented including solar panels capacity, the number of needed solar panels, the price of every solar panel, etc. In order to analyze the two models and compare the results with each other, the optimized model is considered for both aforementioned, hiring the simulation method to the models. Finally, the test results of the simulation and mathematical programming are presented to indicate the advantages and disadvantages of the two models. Besides, the study can guide users whether to utilize the electricity production obtained from the construction of solar power plants out of the city, or from the photovoltaic generations installed on every building. To make the paper clear, the following sections are presented with the aim of focusing on the discussed subjects.

## II. SOLAR ENERGY FOR ELECTRICITY PRODUCTION

Using solar energy as a source of producing electricity has certain advantages and disadvantages. Although there are many advantages of using solar energy- including the decrease of environmental pollution, the cost-effectiveness compared to the projected high cost of oil that is renewable, sustainable, endless, and easy to use [18]. One of the main disadvantages is the initial cost of the equipment used to harness the energy from the sun. It is obvious as the price of solar panels decreases, the use of solar cells to generate electricity will increase [26]. To be efficient in providing a source of electricity, a solar energy installation requires a large area to host the system. This may be a discouragement in areas where space is limited, or expensive. Pollution can be a disadvantage to solar panels since pollution can degrade the efficiency of photovoltaic cells. Clouds also provide the same influence since they can partially block the energy of the sun's rays. This certain disadvantage is more of an issue with older solar components, as newer designs integrate technologies to overcome the worst of these effects.

Also, solar energy is only useful when the sun is shining, so it requires a backup power. During the night, the expensive solar equipment will remain unused despite the fact that the use of solar battery chargers can help to reduce the influence of this drawback; therefore, the location of solar panels can have an effect on the performance. For instance, TEP operates one of the largest solar PV arrays in the United States, a 5-MW system. But over two years of operation, the capacity factor for that generator has averaged 19%, meaning it produced only 19% of its rated capacity most of the time [27]. It should be noted that solar energy as a renewable source of electricity generation has some limitations which need special consideration in order to improve efficiency and reduce costs. In this paper, authors have tried to apply mathematical programming and simulation methods to two models of PV solar power plants and PV generations using the lean manufacturing method for the latter in order to represent the test results for both models.

## III. LEAN MANUFACTURING

Lean manufacturing emphasizes various parameters including those of reducing waste, boosting efficiency, and obtaining continuous improvement. It focuses on minimizing costs, maximizing customer options, speeding up delivery, and increasing the quality of products and services. Proponents of lean manufacturing have identified roughly eight major types of wastes, namely overproduction, waiting for the next process step, unnecessary transport of materials, over-processing of parts, inventories more than minimum amounts, unnecessary movement by employees, production of defective parts, and underutilization of human resources. More broadly, waste can be defined as any activity that does not add value to the product or service.

The guiding principle in lean manufacturing is the elimination of non-value-adding activities through continuous improvement efforts. In general, waste consumes resources but creates no value for customers. Waste reduction is thus essential and can include the elimination of excess usage of utilities and materials. The focus of the multidimensional approach is on cost reduction, by eliminating the non-value-added activities and using tools such as cellular manufacturing. It also focuses on total productive maintenance, production smoothing, setup reduction and the like to omit the waste [28], and [29].

Modeling is considered for two cases in this paper. In the first case, it is made for solar power plants installed out of the city whereas the second is the consideration of lean manufacturing for photovoltaic generations on buildings in urban areas. Therefore, by applying lean manufacturing to the second case electricity transition costs are omitted. Also, waste transfer and additional processes, costs of electricity production and of land needed for solar power plants are subtracted. It is obvious that analyzing the two cases indicates the optimal model.

## IV. STRUCTURE OF THE MODEL

In this paper, we use these formulations to model solar power plants and photovoltaic power generations. In this section, the structure of optimized model with the combination of lean manufacturing approach and nonlinear programming are presented, in the first and the second model, we try to minimize and optimize costs of produced power for solar power plants installed and generated out of the city and photovoltaic power generations on buildings respectively. Finally, in the third model, the number of solar panels used for buildings in the city and solar power plants out of the city accorded with the power production demand are presented and compared.

## V. FIRST PROBLEM FORMULATION

In this model, the power plant is a photovoltaic power plant, and all the solar panels have the same performance.

The first model is presented as the followings:

A. *Indices*

The following indexes are considered:

$j$ : Index of power plants $(j = 1, 2, ..., m)$

$t$ : Index of time periods $(t = 1, 2, ..., T)$

B. *Parameters*

The following parameters are considered:

$C_{jt}$ : Cost per kilowatt-hour power produced by $j$ power plant in period $t$

$V_{jt}$ : Cost per kilowatt-hour power transferred from $j$ power plant in period $t$

$H_{jt}$ : Cost of excess power produced by $j$ the power plant in period $t$

$D_{jt}$ : Power Consumption received from $j$ the power plant (uniform distribution)

$R_j$ : Setup cost for $j$ power plant

$\Delta_{jt}^{max}$ : Maximum capacity $j$ power plant in the period $t$ in per unit

$\Delta_{jt}^{min}$ : Minimum capacity $j$ power plant in the period $t$ per unit.

$F$ : Certain number of power plants

$NPW_{jt}$ : Net present value costs (operational costs) for $j$ the power plant in period $t$

C. *Variables*

Variables are defined as followings:

$Y_j$ : 1 if $j$ the power plant is used, 0 otherwise.

$Z_{jt}$ : Power produced by $j$ power plants per kilowatt-hour in period $t$.

$K_{jt}$ : Energy surplus produced by $j$ power plants in period $t$.

$Z_{optimum}$ : Objective Function of the primary model.

D. *Model*

A nonlinear programming method is utilized in this model to obtain optimal variables.

$$Min Z_{optimum} = \sum_{j=1}^{m} R_j Y_j$$
$$+ \sum_{j=1}^{m} \sum_{t=1}^{T} (NPW_{jt} + V_{jt}) Y_j Z_j (1+i)^{-t} \quad (1)$$
$$+ \sum_{j=1}^{m} \sum_{t=1}^{T} H_{jt} K_{jt} (1+i)^{-t}$$

Subject to:

$$\Delta_{jt}^{min} \leq Y_j Z_{jt} \leq \Delta_{jt}^{max}, \forall j = 1, 2, ..., m, \forall t = 1, 2, ..., T \quad (2)$$

$$K_{jt} = Y_j Z_{jt} - D_{jt}, \forall j = 1, 2, ..., m, \forall t = 1, 2, ..., T \quad (3)$$

$$\sum_{j=1}^{m} Y_j = F, \forall j = 1, 2, ..., m \quad (4)$$

$$Y_j = \{0,1\}, Z_{jt} \geq 0, K_{jt} \in R,$$
$$\forall j = 1, 2, ..., m, \forall t = 1, 2, ..., T \quad (5)$$

In the first part of the objective function, the cost of setting up a Power plant $j$ is considered. In the second part of the objective function, operating costs and transportation costs per *kWh* of electricity during the period $t = 1, 2, ..., T$ for the power plant $j$ are shown. In the third part of the objective function, the cost of additional power generation in the power plant $j$ during $t$ is indicated. The constraint shown as number two is to prevent exceeding the maximum and minimum amount of electricity production while choosing the power plant.

VI. SECOND PROBLEM FORMULATION

This model analyzes photovoltaic power generations installed on the buildings of the city based on a mathematical model. The decision making is presented accorded with the parameters which are vital to the photovoltaic power generation analysis.

A. *Parameters*

The following parameters are considered:

$I$ : The interest rate.

$T$ : Production life (operation period)

$Q$ : Operation cost

$C$ : cost to purchase one solar panel

$A$ : Total consumption criterion quantity

$B$ : Production capacity of each solar panel

$N$ : The number of solar panels needed.

B. *Variables*

Variables are defined as followings:

$Z$ : Electricity produced by each solar panel

$F$ : The objective function of the second model

C. *Model*

This model is for photovoltaic power generations installed on buildings of the city. It is based on the mathematical model [30]:

$$F = \frac{A.C.Q.Z}{B} + \int \sum_{t=1}^{T} \frac{A.Q.Z}{B} (1+I)^{-t} .dZ \quad (6)$$

$$F = \frac{A.C.Q.Z}{B} + \int \frac{A.Q.Z}{B} \left( \frac{1 - \left(\frac{1}{1+I}\right)^T}{I(I+1)} \right) .dZ \quad (7)$$

$$F = \frac{A.C.Q.Z}{B} + \frac{A.Q.Z^2}{2.B} \left( \frac{1 - \left(\frac{1}{1+I}\right)^T}{I(I+1)} \right) \quad (8)$$

$$\frac{\partial F}{\partial Z} = 0 \quad (9)$$

$$\frac{A.C.Q}{B} + \frac{A.Q.Z}{B}\left(\frac{1-\left(\frac{1}{1+I}\right)^T}{I(I+1)}\right) = 0 \quad (10)$$

$$\frac{A.Q.Z}{B}\left(\frac{1-\left(\frac{1}{1+I}\right)^T}{I(I+1)}\right) = -\frac{A.C.Q}{B} \quad (11)$$

$$Z^* = \left|\frac{-C.I.(I+1)}{1-\left(\frac{1}{1+I}\right)^T}\right| \quad (12)$$

The derivative of the model is calculated to obtain the optimal amount of $Z^*$, then by entering the amount, the optimized model is presented:

$$F = \frac{A.C.Q}{B} \cdot \left[\frac{-C.I.(I+1)}{1-\left(\frac{1}{1+I}\right)^T}\right] + \quad (13)$$

$$\frac{A.Q}{2.B} \cdot \left[\frac{-C.I.(I+1)}{1-\left(\frac{1}{1+I}\right)^T}\right]^2 \cdot \left(\frac{1-\left(\frac{1}{1+I}\right)^T}{I(I+1)}\right)$$

$$F = \frac{A.C.Q}{B} \cdot \left[\frac{-C.I.(I+1)}{1-\left(\frac{1}{1+I}\right)^T}\right] \quad (14)$$

$$+ \frac{A.Q.C^2}{2.B} \cdot \left[\frac{I(I+1)}{1-\left(\frac{1}{1+I}\right)^T}\right]$$

$$F = \frac{-A.C^2.Q.I(I+1)}{2.B.\left[1-\left(\frac{1}{1+I}\right)^T\right]} \quad (15)$$

## VII. DEPENDENCE ON THE NUMBER OF SOLAR PANELS WITH POWER PLANTS

In the first model, $z_i$ is calculated by the following formula:

$$z_i = \frac{-\sum_{j=1}^{m}\int_{a}^{x} H_j.K_j.dK_j(1+i)^{-t} - \sum_{j=1}^{m} R_j.Y_j}{\sum_{j=1}^{m} NPW_j.Y_j} \quad (16)$$

In the second model, $Z$ can be calculated as the first model:

$$Z = \frac{-N.C.Q + \sqrt{(N.C.Q)^2 + 2.F.N.Q.\left(\frac{1-\left(\frac{1}{1+I}\right)^T}{I(I+1)}\right)}}{N.Q.\left(\frac{1-\left(\frac{1}{1+I}\right)^T}{I(I+1)}\right)} \quad (17)$$

In order to calculate $N^*$, the two above formulas are set equal to each other as the following:

$$\frac{-N.C.Q + \sqrt{(N.C.Q)^2 + 2.F.N.Q.\left(\frac{1-\left(\frac{1}{1+I}\right)^T}{I(I+1)}\right)}}{N.Q.\left(\frac{1-\left(\frac{1}{1+I}\right)^T}{I(I+1)}\right)} \quad (18)$$

$$= \frac{-\sum_{j=1}^{m}\int_{a}^{x} H_j.K_j.dK_j(1+i)^{-t} - \sum_{j=1}^{m} R_j.Y_j}{\sum_{j=1}^{m} NPW_j.Y_j}$$

Then, $N^*$ is obtained from the above equality. It is conspicuous that by entering the parameters of the third model, the variables are obtained continuously, so when $Z^*$ is entered, $N^*$ is obtained.

## VIII. SIMULATION AND THE RESULTS

In this paper, it is used as a suitable method for the operational analysis of solar power plants and solar power generations since the study of physical systems is very difficult and expensive or even impossible in certain cases. This process was performed through Arena simulation software.

In the first model, the main aim of the simulation model analysis is to evaluate the effects of uncertainties on-demand rates. Since the demand rates are uncertain, they can be characterized by a certain probability distribution. In this section, to investigate the solar power plants by a simulation method.

### A. Results of the first model

The results of the optimization model are presented according to different input data shown on three tables including low, medium, and high demand respectively. In table 1, the input data of the optimization model are entered into the Lingo 8 software and the output is given on the right. The optimum amount of the objective function is obtained from the software $(Z = 43600)$.

Table 1 data of the first model (low demand)

| The output | $K_{jt}$ | 0 | 0 | 0 | 0 | $3.9*10^3$ | $3.8*10^3$ | 0 | 0 |
|---|---|---|---|---|---|---|---|---|---|
| | $Z_{jt}$ | $4*10^3$ | $4.2*10^3$ | $4.3*10^3$ | $4*10^3$ | 0 | 0 | 0 | $4.1*10^3$ |
| | $Y_j$ | 1 | | 1 | | 1 | | 1 | |

| The input | | | | | | | | |
|---|---|---|---|---|---|---|---|---|
| Row | | $R_j$ | $NPW_{jt}$ | $V_{jt}$ | $H_{jt}$ | $\Delta_{jt}^{min}$ | $\Delta_{jt}^{max}$ | $D_{jt}^{min}$ |
| t=1 | j=1 | $5*10^9$ | $3.5*10^4$ | $3.4*10^4$ | $1.2*10^4$ | $3*10^3$ | $4*10^3$ | $4*10^3$ |
| t=2 | j=1 | $5*10^9$ | $4.5*10^4$ | $4*10^4$ | $1.3*10^4$ | $3.5*10^3$ | $4.5*10^3$ | $4.2*10^3$ |
| t=1 | j=2 | $3.5*10^9$ | $4.5*10^4$ | $3.5*10^4$ | $1.8*10^4$ | $3.2*10^3$ | $4*10^3$ | $4.3*10^3$ |
| t=2 | j=2 | $3.5*10^9$ | $2.5*10^4$ | $3*10^4$ | $2*10^4$ | $3.3*10^3$ | $4*10^3$ | $4*10^3$ |
| t=1 | j=3 | $4*10^9$ | $3.2*10^4$ | $2.5*10^4$ | $1.4*10^4$ | $3.1*10^3$ | $4*10^3$ | $3.9*10^3$ |
| t=2 | j=3 | $4*10^9$ | $3*10^4$ | $2.9*10^4$ | $1.5*10^4$ | $3.2*10^3$ | $4*10^3$ | $3.8*10^3$ |
| t=1 | j=4 | $3*10^9$ | $2.8*10^4$ | $4*10^4$ | $9*10^3$ | $3.4*10^3$ | $4*10^3$ | $4*10^3$ |
| t=2 | j=4 | $3*10^9$ | $5*10^4$ | $3.9*10^4$ | $1.5*10^4$ | $3.5*10^3$ | $4.5*10^3$ | $4.1*10^3$ |

Table 2 demonstrates the input and output data of the second sample (medium demand). The output data and the optimum amount of the objective function $(Z=51300)$ are calculated by *Lingo 8* software. (The first 8 columns are the same as table 1)

Table 2 data of the first model (medium demand)

| The input | The output | | |
|---|---|---|---|
| $D_{jt}^{min}$ | $Y_j$ | $Z_{jt}$ | $K_{jt}$ |
| $4.5*10^3$ | 1 | 0 | $5*10^3$ |
| $4.7*10^3$ | | $5.2*10^3$ | 0 |
| $4.8*10^3$ | 1 | $5.3*10^3$ | 0 |
| $4.5*10^3$ | | 0 | $5*10^3$ |
| $4.4*10^3$ | 1 | $5.9*10^3$ | 0 |
| $4.3*10^3$ | | $5.8*10^3$ | 0 |
| $4.5*10^3$ | 0 | $4*10^3$ | $5*10^3$ |
| $4.6*10^3$ | | $4.1*10^3$ | $5.1*10^3$ |

In table 3, the input and output data of the third sample (high demand) are illustrated. As mentioned earlier, the output data and the optimum amount of the objective function $(Z=36000)$ are obtained from *Lingo 8* software. (The first 8 columns are the same as table 1)

Table 3 data of the first model (high demand)

| The input | The output | | |
|---|---|---|---|
| $D_{jt}^{min}$ | $Y_j$ | $Z_{jt}$ | $K_{jt}$ |
| $5*10^3$ | 1 | 0 | $4.5*10^3$ |
| $5.2*10^3$ | | $4.7*10^3$ | $4.5*10^3$ |
| $5.3*10^3$ | 1 | $4.8*10^3$ | 0 |
| $5*10^3$ | | 0 | 0 |
| $4.9*10^3$ | 1 | $5.4*10^3$ | 0 |
| $4.8*10^3$ | | $5.3*10^3$ | 0 |
| $5*10^3$ | 0 | 0 | $4.5*10^3$ |
| $5.1*10^3$ | | 0 | $4.6*10^3$ |

*B. Results of the second model*

In this section, the presented results of the second model according to the data are shown in table 4 hat have five alternatives for calculating an optimization modeling of photovoltaic power generations in this table (4).

Table 1)

Table 4 Parameters of the second model

| | | Alternatives | | | | |
|---|---|---|---|---|---|---|
| | | Korea | China | Taiwan | U.S.A. | Japan |
| input | I | 0.25 | 0.12 | 0.18 | 0.10 | 0.13 |
| | T (month) | 60 | 12 | 48 | 36 | 24 |
| | Q (year/$) | 80 | 10 | 50 | 90 | 50 |
| | C ($) | 410 | 170 | 250 | 390 | 433 |
| | A (W/year) | 456250 | 32850 | 423400 | 279225 | 175200 |
| | B (W) | 250 | 90 | 290 | 255 | 240 |
| output | Z* | 191.23 | 213.24 | 109.66 | 172.50 | 138.71 |
| | F ($) | $5.7*10^9$ | $6.6*10^7$ | $1*10^9$ | $3.3*10^9$ | $1*10^9$ |

The optimum number of solar panels used for the photovoltaic power generations installed on buildings of the city is $1*10^{29}(N*)$ based on the optimum amount of $Z$ obtained from the above model.

## IX. CONCLUSION

In this paper, by using optimization models and simulation methods, an approach is presented including the combination of mathematical and economic models used for distribution systems of electricity transmission networks in solar power plants installed and generated out of the city and photovoltaic power generations on buildings of the city as a lean manufacturing process. The contribution of the paper is that there was no specific published record of comparing the simulation and optimization model applied to both solar power plants and photovoltaic power generations. The two models were compared and the results showed that the optimization and simulation models have practically led us to the same results. It can be concluded that the first and the second models have the same operation, which is really useful for using both methods as reliable techniques of analyzing the operation of solar power plants and photovoltaic power generation utilized in different buildings. It is obvious that the operation analysis obtained from the two models is a procedure used to determine the efficiency. Although different results can be obtained from different situations based on various data of the first and the second models, in this paper, photovoltaic power generations installed on the buildings yield to better results in comparison with solar power plants installed and generated out of the city due to the comparison of the value of $Z$, and $F*$. For instance, according to table 5 and table 8, the value of $Z$ for the solar power plants (Low demand) was $(Z=43600)$ and the value of $F*$ for the photovoltaic power generations in Alt1 (Korea) was $(F^* = 16346.08)$. Therefore, it can be concluded that solar power generations have better economic performance in comparison to solar power plants. Consequently, the main objective of this paper is to present the two mentioned methods as reliable techniques for analyzing and evaluating different solar power plants and photovoltaic power generations for the sake of introducing the optimum model which has the minimum rate of energy waste and the maximum efficiency.